\documentclass[prb,twocolumn,showpacs,floatfix,amsmath,amssymb,superscriptaddress,longbibliography]{revtex4-2}
\usepackage{amsfonts}
\usepackage{stmaryrd}
\usepackage{mathrsfs}
\usepackage{tipa}
\usepackage{amssymb}
\usepackage{txfonts}
\usepackage{graphicx}
\usepackage{dcolumn}
\usepackage{epstopdf}
\usepackage[colorlinks,linkcolor=blue,urlcolor=blue,citecolor=blue]{hyperref}
\usepackage{multirow}
\usepackage{subfigure}
\usepackage{url}
\usepackage{upgreek}
\usepackage[utf8]{inputenc}
\usepackage[english]{babel}
\usepackage{siunitx}
\usepackage{bm}
\newcommand{\im}{\mathrm{i}}
\newcommand{\vc}{\bm}
\newcommand{\mat}{\mathbf}
\newcommand{\real}{\text{Re}}
\newcommand{\imag}{\text{Im}}
\newcommand{\Ham}{\mathcal{H}}
\newcommand{\ti}{\tilde{t}}

\usepackage{xspace}
\usepackage[abbreviations, hyperfirst=true]{glossaries-extra}
\newcommand\createabbrv[3]{%
    \newabbreviation{#1}{#2}{#3}%
    \expandafter\newcommand\csname #2\endcsname{\gls{#1}\xspace}%
}

\usepackage{xcolor}

\begin{document}

\title{Low-energy spin excitations in field-induced phases of the spin-ladder antiferromagnet BiCu$_2$PO$_6$}

\author{Patrick~Pilch}
\affiliation{Department of Physics, TU Dortmund University, 44227 Dortmund, Germany}

\author{Kirill~Amelin}
\affiliation{National Institute of Chemical Physics and Biophysics, 12618 Tallinn, Estonia}

\author{Gary~Schmiedinghoff}
\affiliation{Institute of Software Technology, German Aerospace Center (DLR), 51147 Cologne, Germany}

\author{Anneke~Reinold}
\author{Changqing~Zhu}
\affiliation{Department of Physics, TU Dortmund University, 44227 Dortmund, Germany}

\author{Kirill~Yu.~Povarov}
\affiliation{Dresden High Magnetic Field Laboratory (HLD-EMFL) and W\"urzburg-Dresden Cluster of Excellence ct.qmat, Helmholtz-Zentrum Dresden-Rossendorf (HZDR), 01328 Dresden, Germany}

\author{Sergei~Zvyagin}
\affiliation{Dresden High Magnetic Field Laboratory (HLD-EMFL) and W\"urzburg-Dresden Cluster of Excellence ct.qmat, Helmholtz-Zentrum Dresden-Rossendorf (HZDR), 01328 Dresden, Germany}

\author{Hans~Engelkamp}
\affiliation{High Field Magnet Laboratory (HFML-EMFL), Radboud University, Toernooiveld 7, 6525 ED Nijmegen, The Netherlands}

\author{Yin-Ping~Lan}

\author{Guo-Jiun~Shu}
\affiliation{Department of Materials \& Mineral Resources Engineering, Institute of Mineral Resources Engineering,
National Taipei University of Technology, Taipei 10608, Taiwan}

\author{F.~C.~Chou}
\affiliation{Center for Condensed Matter Sciences, National Taiwan University, Taipei 10617, Taiwan}

\author{Urmas~Nagel}
\author{Toomas~Rõõm}
\affiliation{National Institute of Chemical Physics and Biophysics, 12618 Tallinn, Estonia}

\author{Götz~S.~Uhrig}
\affiliation{Department of Physics, TU Dortmund University, 44227 Dortmund, Germany}

\author{Benedikt~Fauseweh}
\email{benedikt.fauseweh@tu-dortmund.de}
\affiliation{Department of Physics, TU Dortmund University, 44227 Dortmund, Germany}
\affiliation{Institute of Software Technology, German Aerospace Center (DLR), 51147 Cologne, Germany}

\author{Zhe Wang}
\email{zhe.wang@tu-dortmund.de}
\affiliation{Department of Physics, TU Dortmund University, 44227 Dortmund, Germany}

\date{\today}

\begin{abstract}
We report on terahertz spectroscopic measurements of quantum spin dynamics on single crystals of a spin-1/2 frustrated spin-ladder antiferromagnet BiCu$_2$PO$_6$ as a function of temperature, polarization, and applied external magnetic fields.
Spin triplon excitations are observed at zero field and split in applied magnetic fields. 
For magnetic fields applied along the crystallographic $a$ axis, a quantum phase transition at $B_{c1}=21.4$~T is featured by a low-energy excitation mode emerging above $B_{c1}$ which indicates a gap reopening.
For fields along the \textit{b} axis and the \textit{c} axis, different field dependencies are observed for the spin triplon excitations, whereas no low-lying modes could be resolved at field-induced phase transitions. 
We perform a theoretical analysis of the magnetic field dependence of the spin triplon modes by using continuous unitary transformations to determine an effective low energy Hamiltonian.
Through an exhaustive parameter search we find numerically optimized parameters to very well describe the experimentally observed modes, which corroborate the importance of significant magnetic anisotropy in the system.
  
\end{abstract}

\maketitle

\section{Introduction}
Low-dimensional quantum magnets are a captivating material class in condensed matter physics due to their manifestation of quantum fluctuations and/or unconventional ordered spin states due to enhanced quantum effects.
Quantum spin-dimerized antiferromagnets feature a spin-singlet ground state which is separated from gapped $S=1$ spin-triplet excitations - the so-called triplons \cite{sachd90,schmi03c}. 
In a basic scenario, the application of a magnetic field $B$ leads to the splitting of the triplon modes.
The low-lying triplon branch softens until the gap closes at a critical field $B_c$ where the Zeeman energy equals the spin gap value at zero field, leading to a field-induced quantum phase transition.
Depending on the interplay between the repulsive boson interaction arising from the hardcore nature of triplons and the gain in kinetic energy from the Zeeman energy, the field-induced phases can exhibit exotic properties, such as magnon Bose-Einstein condensation (BEC) indicated by staggered magnetization \cite{giamarchi2008bose, ruegg2003bose}, or supersolid states close to fractional magnetization plateaus (see e.g. \cite{Yamamoto2013Magnon, rice2002condense, Matsuda2013Magnetization}).

\begin{figure*}[t!]
\centering
 \includegraphics[width=0.85\linewidth]{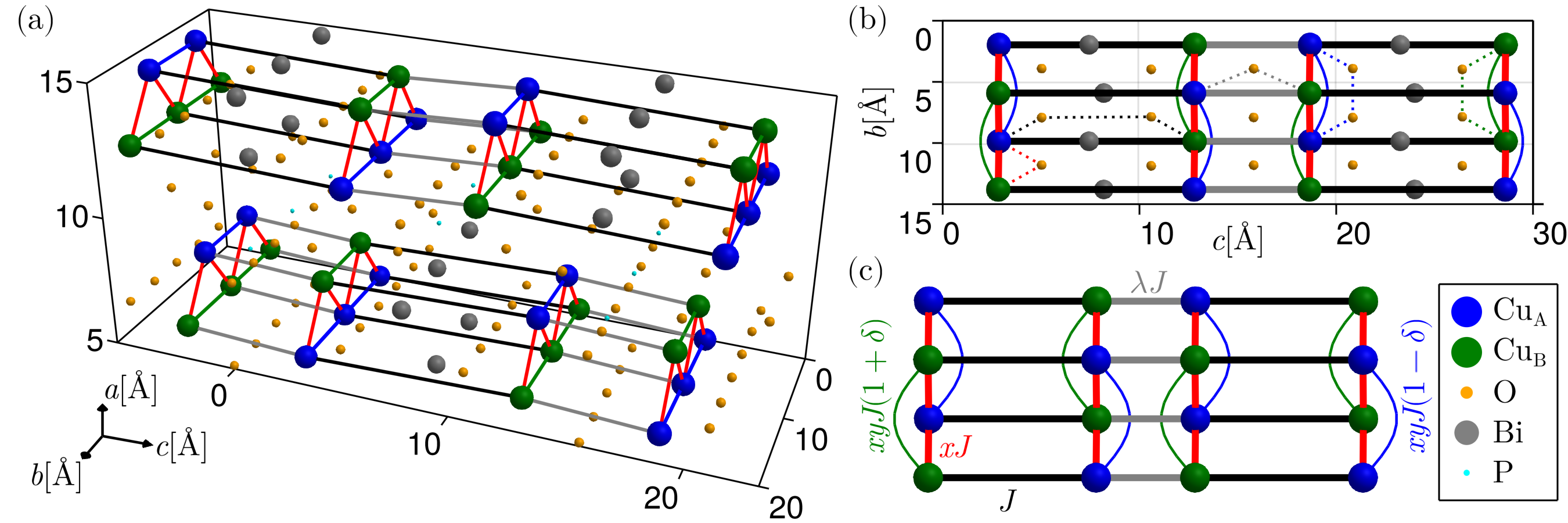}
 \caption{\label{fig:BCPO_model}
Structure for BCPO with colored lines to indicate interactions. The (a) three-dimensional view reveals two-dimensional bilayers of interactions, the latter of which are depicted by straight lines. Panel (b) shows a single bilayer from above, where the dotted lines indicate the oxygen atoms involved in each (super-)exchange interaction. The final, abstracted spin model including the names of the interactions is depicted in (c). Each interacting pair on the same ladder is also subjected to the SAE and DM interactions. }
\end{figure*}

In spin-$1/2$ antiferromagnets, a singlet-triplet gap can arise in even-legged spin ladders \cite{White_RVB94}, spin dimerized systems \cite{Wang11,Wang14}, spin tubes \cite{Garlea2008Excitations, Nishimoto2011Spingap, Gomez2014Magnetization}, or spin chains with alternating exchange interactions \cite{Johnston2000PRB} or frustrated next-nearest neighbor couplings \cite{Haldane_Spon_1982, OKAMOTO1992433}, or exchange isotropy \cite{Wang2018,Wang2024}.
BiCu$_2$PO$_6$ (BCPO), a frustrated two-leg spin ladder with alternating next-nearest-neighbor interactions, is therefore a compelling system for the investigation of magnetic ordering phenomena and quantum phase transitions driven by frustration and quantum fluctuations \cite{Tsirlin2010bridging}. 
The magnetic structure of BiCu$_2$PO$_6$ is presented in Fig.~\ref{fig:BCPO_model} \cite{Tsirlin2010bridging, colmont2018compressibility, hwang2016theory, koteswararo2007spin, mentre2009incommensurate}, where the exchange interactions between Cu$^{2+}$ $S=1/2$ spins with two inequivalent crystallographic sites are indicated by different bond colors.
Spin ladders are running in the \textit{bc} plane along the \textit{b} axis with the intra-rung exchange interaction denoted by $J$. Along the ladder both nearest- and next-nearest neighbor exchanges [see Fig.~\ref{fig:BCPO_model}(c)] are important for understanding the magnetic properties.
Inter-ladder coupling in the \textit{a} direction is smaller by about one order of magnitude.
Dzyaloshinskii-Moriya (DM) interactions are allowed and considerable for certain bonds.
Previous experimental studies of BiCu$_2$PO$_6$ revealed numerous interesting physical phenomena, e.g., a phonon coupling to the spin lattice \cite{Antonakos_2012}, a strong dependence of the ground state on the pressure during crystal growth \cite{colmont2018compressibility}, and a strong phonon contribution to the thermal conductivity alongside a universal spin thermal conductivity contribution \cite{kawamata2018thermal}.
The magnetic properties of BiCu$_2$PO$_6$ have been studied by measuring magnetization \cite{Tsirlin2010bridging, kohama2012anisotropic}, magnetic susceptibility \cite{mentre2006structural, koteswararo2007spin}, heat capacity \cite{koteswararo2007spin, koteswararao2009doping}, Raman scattering \cite{choi2013evidence}, and Knight shift \cite{alexander2010impurity, casola2013DirectObservation}.
Inelastic neutron scattering (INS) experiments \cite{plumb2016quasiparticle, plumb2013incommensurate} have unveiled spin excitations with an incommensurate wave vector in a quantum-disordered low temperature phase. Together with theoretical investigations using quadratic bond-operator theory \cite{plumb2016quasiparticle} and exact diagonalization \cite{Tsirlin2010bridging} these results emphasize the importance of multi-triplon excitations, corroborated by Raman scattering experiments \cite{choi2013evidence}. However, the observed dispersion relation of the magnetic excitations, including the bending of the low-energy branch near the magnetic excitation continuum in the reciprocal space, have not been fully understood despite numerous theoretical attempts \cite{plumb2016quasiparticle, splinter2016minimal,malki2019absence, mueller2023interacting}, which calls for further experimental studies.

A nuclear magnetic resonance (NMR) study in external magnetic fields along the \textit{b} axis revealed deviations from a standard BEC due to an unusual critical exponent, highlighting the significance of anisotropic Dzyaloshinskii-Moriya and the symmetric anisotropic exchange (SAE) interactions \cite{casola2013field}, which is consistent with the findings of previous magnetization measurements \cite{Tsirlin2010bridging}. 
A field-induced intermediate phase with an incommensurate spiral structure \cite{pikulski2020two} was suggested, which was described theoretically by the formation of a soliton lattice, i.e. fractionalized spin-1/2 quantum domain walls \cite{casola2013field, sugimoto2015magnetization}.
In higher fields, the system undergoes a second incommensurate-commensurate (IC-C) phase transition. 
For fields along the \textit{a} or \textit{c} axis, a comprehensive NMR study has not been performed \cite{casola2013field}, but specific heat measurements did not support the formation of a soliton lattice \cite{kohama2014entropy}, where magnetic anisotropy in the \textit{g} tensor or in the Dzyaloshinskii-Moriya interactions could be more influential and bring more complexity to the theoretical analysis \cite{casola2013field}.

In this work we focus on the study of magnetic excitations in BiCu$_2$PO$_6$ by performing high-resolution terahertz (THz) spectroscopy and electron spin resonance (ESR) spectroscopy as a function of temperature and applied high magnetic fields.
The dependence of the magnetic excitations on the applied magnetic field along different crystallographic directions is obtained and quantitatively compared to our theoretical analysis.
In particular, we provide spin dynamic evidence on field-induced quantum phase transition and determine the underlying spin interaction Hamiltonian.    

\section{Experimental details}
Single crystals of BiCu$_2$PO$_6$ were grown by floating-zone method \cite{choi2013evidence}. We prepared two sample orientations: a wedged $a-$cut sample measuring $5\times \SI{5}{\milli\meter^2}$ with an average thickness of $\SI{1}{\milli\meter}$ and a $b-$cut sample of the same dimensions but with a thickness of $\SI{1.1}{\milli\meter}$.
Zero-field polarized THz transmission spectra were acquired using the TeslaFIR spectrometer in a magneto-optical cryostat \cite{kezsmarki2014one}, and differential absorption $\Delta \alpha(T)=\alpha(T)-\alpha(30\,\mathrm{K})$ was calculated using the $T=\SI{30}{\kelvin}$ spectrum as a reference. Similarly, polarized low-field THz spectroscopy experiments ($B\leq\SI{16}{\tesla}$) were conducted with magnetic fields applied using a superconducting magnet in Faraday and Voigt geometry. As a reference, the $\SI{0}{\tesla}$ spectrum was used to calculate the differential absorption $\Delta\alpha(B)=\alpha(B)-\alpha(0\mathrm{T})$. $\alpha(0\mathrm{T})$ was recovered from  a median of all measured  $\Delta\alpha(B)$. This method allows to extract the magnetic-field dependent part of the absorption coefficient. 
Unpolarized high-field ESR experiments were performed at the Dresden High Magnetic Field Laboratory (Hochfeld Magnetlabor-Dresden), using a multi-frequency  ESR spectrometer (similar to that described in Ref.~\cite{ZVYAGIN20041}), in
magnetic fields up to $\SI{60}{\tesla}$.
We used VDI microwave-chain radiation sources (product of Virginia Diodes, Inc., USA) to generate radiation in the sub-THz frequency range. A hot-electron n-InSb bolometer (product of QMC Instruments Ltd., UK), operated at $\SI{4.2}{\kelvin}$, was used as a radiation detector. Additional unpolarized high-field spectroscopy measurements with magnetic fields up to $\SI{28}{\tesla}$ and at a temperature of $\SI{1.4}{\kelvin}$ were performed at the Nijmegen High Field Magnet Laboratory (HFML), using a Bruker IFS 113v Fourier-transform spectrometer. The corresponding differential absorption spectra were obtained by subtracting the average of the absorption spectra of all magnetic fields. 

\section{Experimental results}
\subsection{Temperature dependence}
\begin{figure}[t]
\centering
\includegraphics[width=1\linewidth]{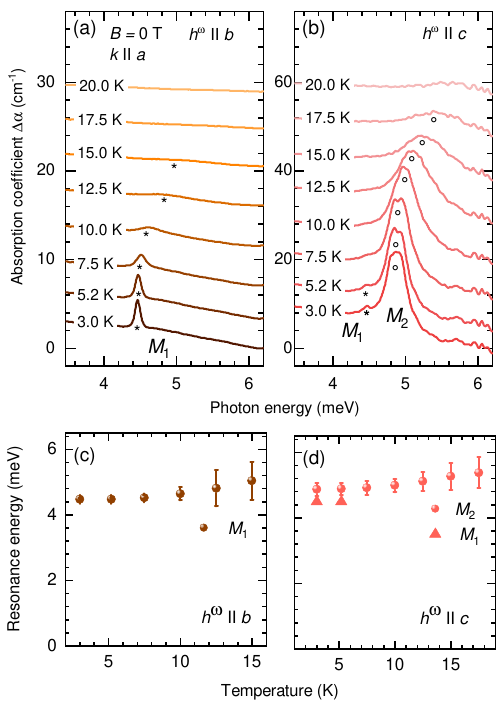}
\caption{Temperature-dependent zero-field excitation spectra measured at $\SI{3}{\kelvin}$ in two polarizations, (a) $h^{\omega}\parallel b$ and (b) $h^{\omega}\parallel c$. The modes $M_1$ and $M_2$ are represented by the asterisks and empty circles, respectively.
(c) and (d) show the resonance energies extracted from the measurements shown in (a) and (b), respectively. The linewidth of the excitations is represented by the vertical bars.}
\label{fig1}
\end{figure}

In zero field, we study the temperature dependence of the magnetic excitations in BiCu$_2$PO$_6$ between $\SI{3}{\kelvin}$ and  $\SI{20}{\kelvin}$, well below the spin gap $\Delta/k_B\approx\SI{34}{\kelvin}$ \cite{mentre2009incommensurate}.
Figure~\ref{fig1} presents the absorption spectra with the THz magnetic field $h^{\omega}$ applied along the crystallographic $b$ axis [Fig.~\ref{fig1}(a)], and along the $c$ axis [Fig.~\ref{fig1}(b)].
At the lowest temperature, two distinct modes emerge at $\SI{4.47}{\milli\electronvolt}$ and $\SI{4.87}{\milli\electronvolt}$, as denoted by $M_1$ and $M_2$, respectively.
A polarization dependence is evidently observed.
The $M_1$ mode exhibits a weaker absorption for $h^{\omega}\parallel c$ than for $h^{\omega}\parallel b$, whereas the $M_2$ mode is observed only for $h^{\omega}\parallel c$ with strong absorption.
The temperature dependence of eigenfrequency and linewidth is summarized in Fig.~\ref{fig1}(c)(d) for the corresponding polarizations.
As the temperature increases, both modes display a thermally induced blueshift and broadening.
Both effects can be explained by the thermal population and hardcore bosonic scattering of triplon modes \cite{PhysRevB.90.024428,PhysRevB.92.214417,PhysRevB.93.241109,PhysRevB.96.115150}.

The observed zero-field modes correspond to the energy of the low-lying excitations as resolved at $Q=(0,0.5,1)$ by INS spectroscopy  \cite{plumb2016quasiparticle}.
Since the THz spectroscopy only probes the $\Gamma$-point, we attribute these modes to zone folding of the triplon branches, which could be due to the inequivalent next nearest neighbor interactions along the \textit{b} axis [see Fig.~\ref{fig:BCPO_model}(c)]. 

\subsection{Magnetic field dependence}

\begin{figure}[t!]
\centering
\includegraphics[width=1\linewidth]{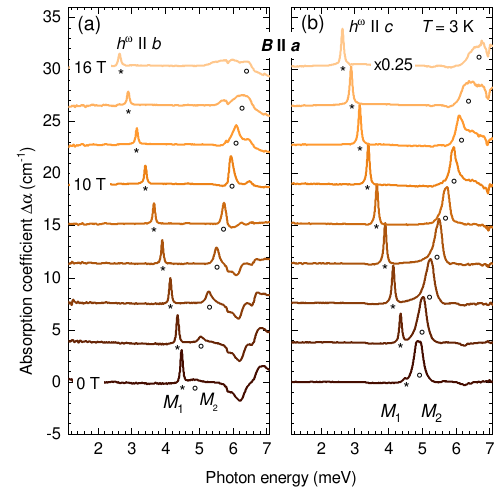}
\caption{Magnetic excitations measured at $\SI{3}{\kelvin}$ in magnetic fields $B\leq\SI{16}{\tesla}$ applied along the crystallographic $a$ axis. Modes are indicated by asterisks ($M_1$) and circles ($M_2$). A constant vertical shift is applied for clarity. The panels show the different polarizations (a) $h^{\omega}\parallel b$ and (b) $h^{\omega}\parallel c$. In panel (b), the spectra are attenuated by a factor of 4 to have comparable peak heights in both panels.}
\label{fig2}
\end{figure}

Figure~\ref{fig2} presents the results for $B \parallel a$ in Faraday configuration. Due to the considerably higher absorption in $h^{\omega}\parallel c$ polarization, the intensity of spectra was multiplied by 0.25 in this polarization. 
We observe distinct behaviors for the two modes, $M_1$ and $M_2$. 
With increasing field $M_1$ softens, while $M_2$ hardens.
A polarization dependence is evident in the strength of the excitations. For $h^{\omega}\parallel b$, $M_1$ weakens and $M_2$ strengthens with the increase of $B$ below $\SI{10}{\tesla}$. Conversely, for $h^{\omega}\parallel c$ the trend is reversed.
At higher fields, $M_2$ broadens indicating a possible shift into a higher-energy phonon mode.
No additional spin excitations were observed at higher energies due to strong absorption of phonon modes above $\SI{6.5}{\milli\electronvolt}$.

\begin{figure}[t]
\centering
\includegraphics[width=1\linewidth]{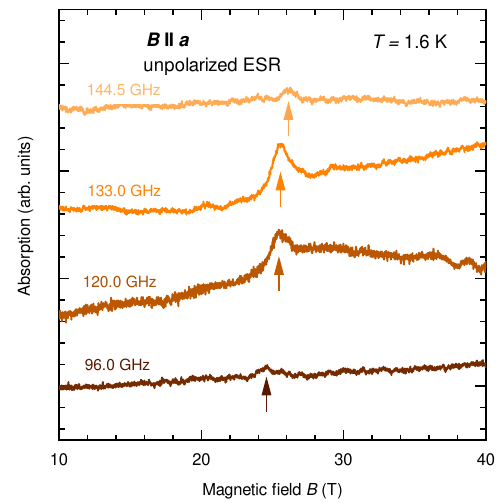}
\caption{Unpolarized high-field ESR measurements performed in pulsed magnetic fields at $\SI{1.6}{\kelvin}$. The absorption curves are shifted proportionally to the photon energy. Arrows point to the absorption peaks.}
\label{fig3}
\end{figure}

To study the spin dynamics when the field is tuned across the phase transition \cite{kohama2014entropy}, we measured ESR transmission spectra of the sample in the same orientation in pulsed magnetic fields, which are presented in
Fig.~\ref{fig3} for different frequencies from 96.0 to 144.5~GHz (corresponding to $\SI{0.4}{\milli\electronvolt}-\SI{0.6}{\milli\electronvolt}$ in photon energy).
As indicated by the arrows, a field-dependent excitation is observed, whose energy increases linearly with increasing field strengths [see Fig.~\ref{fig7}(a)].
However, this mode is detectable only within a narrow frequency range and emerges only above the critical field $B_{c1}$, which is characteristic for the field-induced phase. 
Although formation of a soliton lattice phase was suggested in a field-induced phase for $B \parallel b$ \cite{casola2013field}, our observed mode for $B \parallel a$ is not necessarily related to soliton formation.
This is not only because a soliton lattice may not be formed for $B \parallel a$ \cite{kohama2014entropy}, but also because we cannot resolve the low-lying excitations in the high-field phase for $B \parallel b$.

\begin{figure}[t]
\centering
\includegraphics[width=1\linewidth]{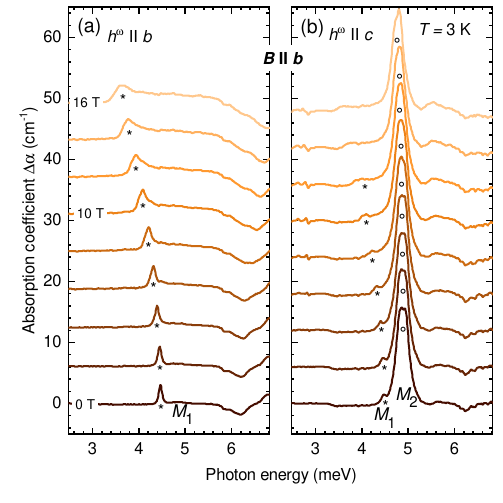}
\caption{Magnetic excitations measured at $\SI{3}{\kelvin}$ in magnetic fields $B\leq\SI{16}{\tesla}$ applied along the crystallographic $b$ axis. Modes are indicated by asterisks ($M_1$) and circles ($M_2$). A constant vertical shift is applied for clarity. The panels show two different polarizations, (a) $h^{\omega}\parallel b$ and (b) $h^{\omega}\parallel c$.}
\label{fig4}
\end{figure}

\begin{figure}[t]
\centering
\includegraphics[width=1\linewidth]{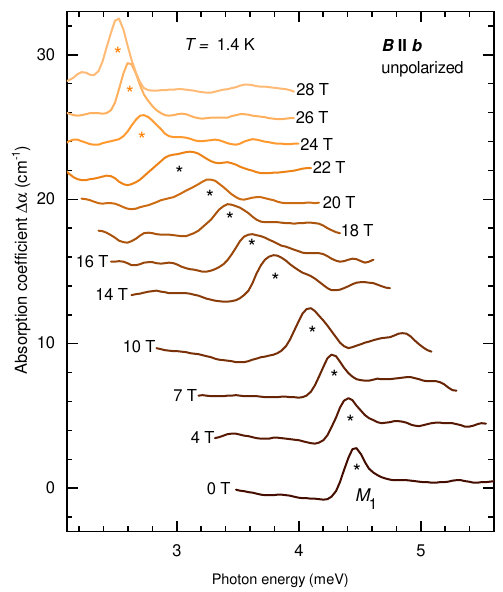}
\caption{Unpolarized high-field FTIR measurements taken at $\SI{1.4}{\kelvin}$ with external magnetic fields $B\leq\SI{28}{\tesla}$ applied along the crystallographic $b$ axis. Asterisks indicate the excitation mode $M_1$ and their color changes above the critical field $B_{c1}\approx\SI{22}{\kelvin}$. The spectra are shifted upwards proportional to the magnetic field strengths.}
\label{fig5}
\end{figure}

\begin{figure}[t]
\centering
\includegraphics[width=1\linewidth]{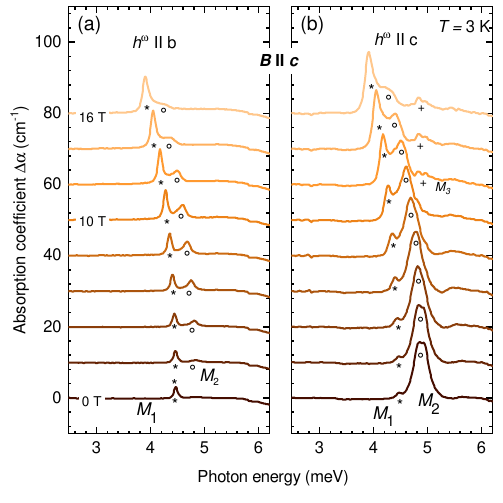}
\caption{Magnetic excitations measured at $\SI{3}{\kelvin}$ in magnetic fields $B\leq\SI{16}{\tesla}$ applied along the crystallographic $c$ axis. Modes are indicated by asterisks ($M_1$), circles ($M_2$) and pluses ($M_3$). A constant vertical shift is applied for clarity. The panels show the different polarizations (a) $h^{\omega}\parallel b$ and (b) $h^{\omega}\parallel c$.}
\label{fig6}
\end{figure}

\begin{figure*}[t]
\centering
\includegraphics[width=0.9\linewidth]{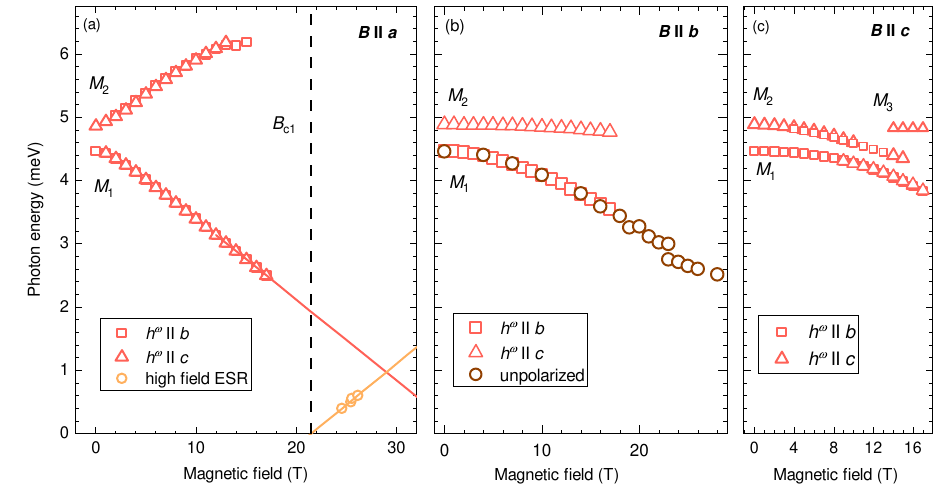}
\caption[width=2\linewidth]{Field-dependent excitations. (a) Modes observed for external magnetic fields applied along the crystallographic $a$ axis. Polarized low-field measurements (red) were supplemented by unpolarized high-field measurements in a pulsed magnet (yellow). (b) Magnetic excitations for magnetic fields applied along the $b$ axis. Polarized low-field measurements (red) and unpolarized high-field measurements (brown) are shown. (c) Spin excitations measured for magnetic fields applied along the $c$ axis in polarized low fields.}
\label{fig7}
\end{figure*}

Figure~\ref{fig4} presents the absorption spectra of the low-field measurements in Voigt geometry for $B \parallel b$. We observe also an evident polarization dependence.
Whereas for $h^{\omega}\parallel b$ the excitation $M_1$ is more pronounced, $M_2$ is substantially enhanced for $h^{\omega}\parallel c$.
We observe a strong field dependence for $M_1$ which is softening with increasing fields. 
The increase of its linewidth indicates a shorter lifetime approaching the critical field $B_{c1}$.
$M_2$ is field-independent for $B \parallel b$ in clear contrast to its hardening behavior for $B \parallel a$. Since $M_2$ is magnetic field-independent in $h^{\omega}\parallel c$, the spectra are calculated from the measured $\Delta\alpha(B)$  using the 3\,K zero field spectrum obtained from zero-field $\Delta\alpha(T)$ spectra.

To track the modes at higher fields and observe their characteristics at field-induced phase transitions, we performed measurements in static high magnetic fields.
As shown in Fig.~\ref{fig5}, only the field-dependent $M_1$ mode was detected, while the field-independent mode $M_2$ was not sensitively resolved by this experimental technique. 
At lower fields, the resonance energies of $M_1$ agree with the results in polarized spectroscopy. 
As the field increases, the softening of the mode becomes more pronounced. Approaching higher field strengths, the mode broadens and strengthens. 
Above the critical field at $\SI{24}{\tesla}$, the mode shifts continuously to lower energies while the softening weakens. 
At the same time, the linewidth decreases with increasing field.
This indicates enhanced fluctuations at the critical field as is inline with approaching a continuous quantum phase transition.

For fields applied along the $c$ axis, i.e. within the ladder plane (see Fig.~\ref{fig:BCPO_model}), the excitation spectra are presented in Fig.~\ref{fig6}(a)(b) for two polarizations $h^\omega\parallel b$ and $h^\omega\parallel c$, respectively.
At zero field the same two modes $M_1$ and $M_2$ are observed. However, their magnetic field dependence for this orientation is different in comparison to the other field orientations. 
While for $h^{\omega}\parallel b$ [Fig.~\ref{fig6}(a)] the mode $M_1$ is more pronounced, we can see a stronger mode $M_2$ for $h^{\omega}\parallel c$ [Fig.~\ref{fig6}(b)]. 
Moreover, another mode $M_3$ appears above $\SI{10}{\tesla}$. With increasing field the $M_1$ and $M_2$ modes soften slightly, whereas the $M_3$ frequency is nearly field-independent.
The peak absorption of the modes varies in different magnetic fields. The $M_1$ mode becomes stronger with increasing magnetic field for both polarizations. 
For $h^\omega \parallel b$, the $M_2$ absorption increases first until $\SI{10}{\tesla}$ and then decreases, while for $h^\omega \parallel c$ its intensity decreases monotonically with increasing field strengths. 
The intensity of the $M_3$ mode increases slightly with field.

We summarize all the observed spin excitations by plotting their eigenfrequencies as a function of magnetic field in Fig.~\ref{fig7} for three magnetic field orientations. 
For $B \parallel a$, we extrapolated the high-field data points of the $M_1$ mode by a linear fit, which yields a critical field of $\SI{36.6}{\tesla}$ and \textit{g}~$\approx 2.20$ that is in good agreement with previous measurements \cite{Tsirlin2010bridging}.
The high-field ESR mode has a much lower energy (see also Fig.~\ref{fig3}). A linear fit to its field dependence yields \textit{g}~$\approx 2.24$ and a critical field of $B_{c1}\approx\SI{21.4}{\tesla}$.
This critical field $B_{c1}$ corresponds to a field-induced phase transition because the spin gap closes.
Its value is consistent with values obtained from specific heat and magnetization measurements \cite{kohama2014entropy, Tsirlin2010bridging}.
Another interesting feature is the crossing of both extrapolations at $B\approx\SI{29.0}{\tesla}$ which corresponds to a critical field for a field-induced first order phase transition \cite{kohama2014entropy}.
For $B \parallel b$, $M_2$ exhibits a slight decrease with increasing field, while the softening of the $M_1$ mode is less pronounced in comparison to $B \parallel c$.
$M_1$ can be followed in the unpolarized measurements even above the critical field $B_{c1}=\SI{23}{\tesla}$  \cite{kohama2014entropy}. 
For $B \parallel c$, the field dependent softening of $M_1$ is even weaker, while the softening of the $M_2$ mode is more evident. $M_3$, resolved above $\SI{12}{\tesla}$, is nearly independent on the field.
Overall, the observed magnetic excitations clearly exhibit different field dependencies for different field orientations, which clearly indicates the presence of magnetic anisotropies. This can be ascribed to anisotropic \textit{g} factors as well as bond-dependent DM interactions, which both will be modelled in detail in the next section.

\section{Theoretical analysis}
In order to describe the magnetic dispersions theoretically, 
we set up an effective model in second quantization of 
the triplons. We employ the approach of continuous unitary transformations (CUT) 
as laid out generally in Refs.\ \cite{knett01b,knett03a} and specifically in
Refs.\ \cite{Krull2012,splinter2016minimal, mueller2023interacting}.
Here, we outline the method in broader strokes and refer the reader to the original work
for more elaborate explanations.

Initially, the system is described by the Hamiltonian
\begin{subequations}\label{eq:H}
\begin{align}
  & \Ham = \Ham_\text{ladder} + \Ham_\text{alter} + \Ham_\text{inter-ladder} + \Ham_{\text{SOC}} + \Ham_\text{magn} ~, \\
& \Ham_\text{ladder} (J,x,y) = J \sum_{rd}  \bigg[ \vc{S}_{r\text{L}d}   \vc{S}_{r\text{R}d} + x \sum_s \vc{S}_{rsd} \vc{S}_{r+1,sd}   \nonumber \\
& \quad  \quad \quad \quad \quad \quad \quad \quad \quad + xy \sum_s \vc{S}_{rsd} \vc{S}_{r+2,sd} \bigg] ~, \label{eq:H_ladder} \\
&   \Ham_\text{alter}(J, x, y,\delta) = Jxy\delta \sum_{rsd}  (-1)^r \vc{S}_{rsd} \vc{S}_{r+2,sd} ~, \label{eq:H_alter} \\
&     \Ham_\text{inter-ladder}(J, \lambda) = \lambda J \sum_{rd} 
    \vc{S}_{rRd} \vc{S}_{rL,d+1} ~, \label{eq:H_interladder} \\
&    \Ham_{\text{SOC}}(\mat{D}, \mat{\Gamma})  =
    \sum_{\langle rsd, r's'd' \rangle} \left[ D^{\phantom{A}}_{rsd,r's'd'} (\vc{S}_{rsd} \times \vc{S}_{r's'd'}) \right. \nonumber \\
&   \quad \quad \quad \quad \quad \quad \quad \quad \quad  \quad    + \sum_{\alpha \beta} \left. \Gamma_{rsd, r's'd'}^{\alpha\beta} S_{rsd}^\alpha S_{r's'd'}^\beta  \right] ~, \label{eq:H_DM} \\
&   \Ham_\text{magn}(\vc{g}) = - \mu_\text{B} \sum_\alpha g^\alpha B^\alpha \sum_{rsd} S_{rsd}^\alpha ~. \label{eq:H_magn}
\end{align}      
\end{subequations}

Figure~\ref{fig:BCPO_model} visualizes the interactions
both in the three-dimensional crystal
as well as in the two-dimensional theoretical model. 
Each vector operator $\vc{S}_{rsd}$
describes a spin on rung $r\in\mathbb{Z}$ and leg
on the left or right-hand side
$s\in\{L,R\}$
of the ladder $d\in\mathbb{Z}$, see Fig.\ \ref{fig:BCPO_model}(c).
The Hamiltonian $\Ham_\text{ladder}$
describes an isotropic Heisenberg spin ladder
with rung couplings $J$.
The energy $J$ sets the dominant energy scale.
The nearest-neighbor (NN) coupling along the legs has strength $xJ$
and the next-nearest-neighbor (NNN) coupling along the legs has strength $xyJ(1\pm\delta)$.
So $x$ describes the relative strength of the NN coupling on the legs to the rung coupling
and $y$ the relative strength of the NNN coupling to the NN  coupling, both on the legs.
The alternation $\pm xyJ\delta$
is described by $\Ham_\text{alter}$.
The Heisenberg coupling of strength $\lambda J$ between 
adjacent ladders is captured by $\Ham_\text{inter-ladder}$.

The Hamiltonian $\Ham_{\text{SOC}}$
captures the effects of the relativistic spin-orbit coupling,
implying the antisymmetric, anisotropic DM interactions $\mat{D}$
and the symmetric anisotropic exchange $\mat{\Gamma}$ \cite{Dzyaloshinsky1958,Moriya1960,Moriya1960a}.
It has been shown that one should consider both the antisymmetric and symmetric anisotropic
terms simultaneously \cite{shekh92}; in leading order they are linked by
\begin{equation}
\label{dm-sae}
    \Gamma^{\alpha\beta}_n =  \frac{D_n^\alpha D_n^\beta}{2J_n} - \frac{\delta_{\alpha\beta} \vc{D}_n^2}{6J_n}
\end{equation}
with $\vc{D}_0 = (0,D_0^b,0)$, $\vc{D}_1 = (D_1^a,D_1^b,D_1^c)$ and $\vc{D}_2 = (D_2^a,0,D_2^c)$.
The index $n$ stands for the rung distance between the interaction partners,
i.e., $J_0 = J$, $J_1 = xJ$, and $J_2=xyJ$.
Equation \eqref{dm-sae} ensures that the SAE does not comprise an isotropic contribution. 
Thus, we only fit the DM terms because  SAE terms are then fixed as well.

The $S_{rsd}^\alpha$ are the components of $\vc{S}_{rsd}$
with $\alpha\in\{a,b,c\}$.
Finally, $\Ham_\text{magn}$ consists of the Zeeman energy with the magnetic field $\vc{B}$, Bohr magneton $\mu_\text{B}$ and the anisotropic $\vc{g}$ tensors.
The model parameters determining  each term are indicated in brackets $\Ham(\cdot)$; they
can be tuned for fitting to the experimental data.

The (approximate) diagonalization of Hamiltonian \eqref{eq:H}
is done in the following steps:
\begin{enumerate}
    \item using a CUT, an effective, bilinear model
    of the single ladder \eqref{eq:H_ladder}
    is computed in terms of the triplon operators.
	In parallel, the operator $S_{rld}^c$
	is also mapped to its leading expression in triplon language,
    \item all other Hamiltonian terms are analytically expressed
    by the effective triplon operators from the previous step,
    \item the eigenvalues are found via a Bogoliubov transformation,
    i.e., by numerically diagonalizing
    the commutation matrix of
    the Hamiltonian and the effective creation and annihilation operators.
\end{enumerate}

\subsection{Computing the Effective Single-Ladder Model}
To obtain an effective model of the coupled spin ladders,
it is convenient to start from the isotropic single-ladder model
$\Ham_\text{ladder} (J,x,y)$ for $J,x,y\geq 0$ and reaching
a description in terms of triplons by CUT.
Note that the antiferromagnetic NN and NNN couplings lead to a frustrated system, i.e.,
frustration is present for $x,y > 0$.
First, we describe the case of separated dimers ($x=y=0$),
for which the elementary excitations are completely local triplets, i.e.,
no distinction between local triplets and the more distributed triplons is necessary.
The expansion in $x$ at finite value $y$ corresponds to
an expansion in the range of the effective couplings
leading to a gradual distribution of the triplons 
over more and more rungs: they become smeared out.
This expansion is stable even for $x, y\gtrapprox 1$,
since the triplons become distributed over more and more rungs, but
still on a finite number of them. They do not delocalize completely
\cite{Krull2012, Schmiedinghoff22_Triplons}.

To express \eqref{eq:H_ladder} in triplon language,
the spin operators are transformed
into triplet operators $\ti_{rd}^\alpha$
of flavor $\alpha \in \{a,b,c\}$ at rung $r$ on ladder $d$
by substituting

\begin{subequations} \label{eq:triplon_ops}
\begin{align}
    2 S_{rLd}^\alpha &= + \ti_{rd}^\alpha + \ti_{rd}^{\alpha \dagger} - \im \sum_{\beta\gamma} \epsilon_{\alpha\beta\gamma} \ti_{rd}^{\beta\dagger} \ti_{rd}^\gamma ~,\\
    2 S_{rRd}^\alpha &= - \ti_{rd}^\alpha - \ti_{rd}^{\alpha \dagger} - \im \sum_{\beta\gamma} \epsilon_{\alpha\beta\gamma} \ti_{rd}^{\beta\dagger} \ti_{rd}^\gamma ~,
\end{align}
\end{subequations}
similar to what is done with bond operators in Ref.\ \cite{sachd90}. 
Note that there can be either none or at most one triplet per rung so that these excitations are hardcore bosons.

The resulting triplet Hamiltonian is systematically mapped to an effective triplon model
\begin{equation}
    \Ham_\text{ladder}^\text{eff}(J,x,y) = J \sum_{kl\alpha} \omega_k(x,y) t_{k}^{\alpha\dagger} t_{k}^{\alpha}
\end{equation}
with triplon operators $t_{k}^{\alpha}$ of flavor $\alpha \in \{a,b,c\}$
and momentum $k$ in $y$-direction (the direction of the ladders)
with single-triplon dispersion $\omega_k(x,y)$. This is achieved by {CUT}s, i.e., a controlled change of basis.
In parallel, the single-spin operator in momentum space $S_{ksd}^c$ is subjected to the
same change of basis, yielding
\begin{equation} \label{eq:S_eff}
    S_{ksd}^{\alpha, \text{eff}}(x,y) = (-1)^s a_k(x,y) (t_{kd}^{\alpha\dagger} + t_{-k,d}^{\alpha}) + \hdots ~.
\end{equation}
This representation is used to express all the parts of the full Hamiltonian 
\eqref{eq:H} which do not belong to the isolated isotropic spin ladders.
All terms in $S_{ksd}^c$ which are nonlinear in the triplon creation and annihilation operators are 
indicated by the three dots in \eqref{eq:S_eff}. Since the other parts of the Hamiltonian
are small relative to the dominant couplings in the isolated isotropic spin ladder
it is justified to neglect these nonlinear terms and to treat the triplons
henceforth as standard bosonic operators, i.e., their hardcore property
is neglected.

The basis change is performed
using a CUT \cite{Kehrein2006, Schmiedinghoff22_Triplons, Schmiedinghoff22_CST},
specifically the deepCUT scheme \cite{Krull2012} which includes infinite
powers of $x$ up to a certain range of interactions.
Finally, all expressions are Fourier transformed in ladder direction $b$.
In the deepCUT, both the Hamiltonian and the observable $S_{rsd}^z$
are expanded in a growing basis of triplon operator monomials
and a set of differential equations is set up which describe the
the continuous basis change. The supplementary information
of Ref.\ \cite{Schmiedinghoff22_Triplons}
describes the general CUT workflow for a spin ladder with $y=0$
in more detail and Ref.\ \cite{Krull2012} provides the specifics of the deepCUT scheme.
Note that the deepCUT needs to be performed
for a fixed parameter pair $(x,y)$ and yields numerical values for
$\omega_k(x,y)$ and $a_k(x,y)$.

The other terms
in \eqref{eq:H} are taken into account
by using the transformed spin observable \eqref{eq:S_eff}
and performing an additional
Fourier transformation in $c$ direction, i.e., perpendicular to
the spin-ladder direction. This leads to a total momentum vector 
$\vc{k} = k \vc{e}_b + l \vc{e}_c = (k,l) $.
The full effective Hamiltonian reads
\begin{widetext}
\begin{subequations} \label{eq:H_eff_allterms}
\begin{align}
    \Ham^\text{eff} &= \Ham_\text{ladder}^\text{eff} + \Ham_\text{alter}^\text{eff} + \Ham_\text{inter-ladder}^\text{eff} + \Ham_{\text{SOC}}^\text{eff} + \Ham_\text{magn}^\text{eff} \label{eq:H_eff} ~, \\
    \Ham_\text{ladder}^\text{eff}(J) &= J \sum_{\vc{k}\alpha} \omega_k t_{\vc{k}}^{\alpha\dagger} t_{\vc{k}}^{\alpha} ~, \\
    \Ham_\text{alter}^\text{eff}(J, \delta) &= 2Jxy\delta \sum_{\vc{k}\alpha} a_k a_{k+\pi}\cos(2k)
    \left( t_{\vc{k}}^{\alpha\dagger}t_{-\vc{k}-(\pi,0)}^{\alpha\dagger}
    + 2 t_{\vc{k}}^{\alpha\dagger}t_{\vc{k}+(\pi,0)}^\alpha
    + t_{\vc{k}}^\alpha t_{-\vc{k}-(\pi,0)}^\alpha \right) ~, \\
    \Ham_\text{inter-ladder}^\text{eff}(J,\lambda) &= 
    -\lambda J \sum_{\vc{k}\alpha} a^2_k \cos(2 l) \bigg( t_{\vc{k}}^{\alpha\dagger} + t_{-\vc{k}}^\alpha \bigg) \bigg( t_{\vc{k}}^\alpha + t_{-\vc{k}}^{\alpha\dagger} \bigg) ~, \\
    \Ham_\text{magn}^\text{eff}(\vc{g}) &= - \im \mu_\text{B} \sum_{\alpha} g^\alpha B^\alpha \sum_{\vc{k}\beta\gamma} \epsilon_{\alpha\beta\gamma} t_{\vc{k}}^{\beta\dagger} t_{\vc{k}}^\gamma
\end{align}
\end{subequations}

with effective DM and SAE interactions

\begin{align} 
    \Ham_{\text{SOC}}^\text{eff}(D_0^b,\vc{D}_1, D_2^c) =
    &+ 4 D_1^c \im \sum_{\vc{k}} a_k^2 \sin(k) \left[ 
    t_{\vc{k}}^{a\dagger} \bigg( t_{(-k,l)}^{b\dagger} + t_{\vc{k}}^b \bigg)
    - \text{H.c.}
    \right] \nonumber \\
    &+ 4 D_2^c \im \sum_{\vc{k}} a_k a_{k+\pi} \sin(2k) \left[ 
    t_{\vc{k}}^{a\dagger} \bigg( t_{(-k-\pi,l)}^{b\dagger} + t_{(k+\pi,l)}^b \bigg)
    - \text{H.c.}
    \right]  \nonumber \\
    &+ \sum_{\vc{k}\alpha} a_k^2 \bigg( 2\Gamma_2^{\alpha\alpha} \cos(2k) + 2\Gamma_1^{\alpha\alpha} \cos(k)-\Gamma_0^{\alpha\alpha} \bigg)
    \left( t_{\vc{k}}^{\alpha\dagger}t_{-\vc{k}}^{\alpha\dagger}
    + 2 t_{\vc{k}}^{\alpha\dagger}t_{\vc{k}}^\alpha
    + t_{\vc{k}}^\alpha t_{-\vc{k}}^\alpha \right)  \nonumber \\
    &+ 2 \sum_{\vc{k}} a_k a_{k+\pi} \left[
    \bigg( \Gamma_1^{ba}e^{-\im k}-\Gamma_1^{ab}e^{\im k} \bigg)
    \bigg( t_{-\vc{k}-(\pi,0)}^{b\dagger} + t_{\vc{k}+(\pi,0)}^b \bigg)
    + \text{H.c.}
    \right] ~. \label{eq:H_eff_SOC}
\end{align}
\end{widetext}
Ref.\ \cite{splinter2016minimal} provides a full symmetry analysis
for the DM and SAE terms.

All remaining model parameters
in \eqref{eq:H_eff_allterms} and \eqref{eq:H_eff_SOC}
that can be easily tuned
are indicated in brackets $\Ham(\cdot)$.
The parameters $(x,y)$ are not listed explicitly,
since changing them requires an additional deepCUT
to obtain updated $\omega_k(x,y)$ and $a_k(x,y)$.
We point out, however, that we tested multiple
pairs $(x,y)$ during fitting
and chose the best result, which turned out to be at $(x,y)=(1.2, 0.85)$ in accordance with Refs. \cite{splinter2016minimal,malki2019absence, mueller2023interacting}.

\subsection{Diagonalization of the Full Effective Hamiltonian}

The eigenenergies of the effective Hamiltonian \eqref{eq:H_eff}
are determined by a generalized Bogoliubov transformation \cite{blaiz86}. 
This is possible if the triplons
are considered as usual bosons without further constraints.
This is justified if the non-diagonal terms
to be taken into account by the Bogoliubov transformation
are small relative to the diagonal ones.
To find the appropriate transformation,
the commutator
\begin{equation}
    [\Ham, v] = w, \quad v = \sum_n v_n O_n, \quad w = \sum_n w_n O_n
\end{equation}
is set up,
where $\{O_n\}$ is a minimal set of basis operators
such that the commutation equations are closed.
Then
the action of the commutation with the Hamiltonian can be
described by a commutation or dynamic matrix $\mat{M}$ \cite{splinter2016minimal,mulle21}
\begin{equation}
[\Ham, O_n] = \sum_m M_{nm} O_m
\end{equation}
which allows one to compute the vectors of the coefficients according to
\begin{equation}
    \mat{M}\vc{v}=\vc{w}~.
\end{equation}
The eigenenergies are found
by diagonalizing the matrix $\mat{M}$.
They appear in pairs $(\omega,-\omega)$ with $\omega>0$ since we always include both, creation 
and annihilation operators.
The eigenvectors provide further information
about which operators contribute
to which eigenmodes.

The minimal basis set for the full Hamiltonian
is $\{O_n\} = \{t_{\vc{k}}^{\alpha\dagger}, t_{\vc{k}+(\pi,0)}^{\alpha\dagger}, t_{-\vc{k}}^\alpha, t_{-\vc{k}-(\pi,0)}^\alpha \}$
with $\alpha \in \{a,b,c\}$.
For a given set of parameters $(x, y, \vc{\theta})$ with $\vc{\theta}=( J, \lambda, D_0^b, \vc{D}_1, D_2^a, D_2^c, \vc{g}, \delta)$,
a 12$\times$12 matrix needs to be diagonalized
for each data point $(\vc{k}, \vc{B})$ of interest, i.e., for each combination of momentum and magnetic field.

The full derivation of the matrix $\mat{M}$
is explained in Refs.\ \cite{splinter2016minimal,mulle21}.
We use
\begin{widetext}
\begin{subequations}
\begin{align}
    \mat{M}(\vc{k}, \vc{B}) &=
    \begin{pmatrix}
        \mat{M}^{aa} & \mat{M}^{ab} & \mat{0}_{4\times4} \\
        \mat{M}^{ba} & \mat{M}^{bb} & \mat{0}_{4\times4} \\
        \mat{0}_{4\times4} & \mat{0}_{4\times4} & \mat{M}^{cc}
    \end{pmatrix}
    + \begin{pmatrix}
        \mat{0}_{4\times4} & \mat{H}^{ab} & \mat{H}^{ac} \\
        \mat{H}^{ba} & \mat{0}_{4\times4} & \mat{H}^{bc} \\
        \mat{H}^{ca} & \mat{H}^{cb} & \mat{0}_{4\times4}
    \end{pmatrix} ~, \label{eq:M_full} \\
    \mat{H}^{\alpha \beta} &= - \im \mu_\mathrm{B} \mat{1}_{4\times4} \sum_\alpha \epsilon_{\alpha\beta\gamma} B^\gamma  g^\gamma ~, \label{eq:M_magn} \\
    \mat{M}^{\alpha\alpha} &=
    \begin{pmatrix}
        \omega_{\vc{k}} + A_{\vc{k}}^{\alpha} & J_{2,k} & -A_{\vc{k}}^\alpha & -J_{2,k} \\
        J_{2,k} & \omega_{\vc{k}+(\pi,0)} + A_{\vc{k}+(\pi,0)}^{\alpha} & -J_{2,k} & - A_{\vc{k}+(\pi,0)}^\alpha \\
        A_{\vc{k}}^{\alpha} & J_{2,k} & -\omega_{\vc{k}} - A_{\vc{k}}^{\alpha} & -J_{2,k} \\
        J_{2,k} & A_{\vc{k}+(\pi,0)}^{\alpha} & -J_{2,k} & -\omega_{\vc{k}+(\pi,0)} - A_{\vc{k}+(\pi,0)}^{\alpha}
    \end{pmatrix} ~, \\
    \mat{M}^{ab} &=
    \begin{pmatrix}
        \mat{C}_k & \mat{C}_k \\
        \mat{C}_k & \mat{C}_k
    \end{pmatrix} ~, \quad
    \mat{M}^{ba} =
    \begin{pmatrix}
        -\mat{C}_k^\mathrm{T} &  \mat{C}_k^\mathrm{T} \\
         \mat{C}_k^\mathrm{T} & -\mat{C}_k^\mathrm{T}
    \end{pmatrix} ~, \quad
    \mat{C}_k =
    \begin{pmatrix}
        \im C_k & \im F_k^{-+} \\
        \im F_k^{++} & \im C_{k+\pi}
    \end{pmatrix} ~, \label{eq:M_ab}
\end{align}
\end{subequations}
\end{widetext}

with
\begin{subequations}
\begin{align}
    A_{k,l}^\alpha &= 2a_k^2
    \left[ -\lambda J\cos(2l) -\Gamma_0^{\alpha\alpha} +2\Gamma_1^{\alpha\alpha}\cos(k) + 2\Gamma_2^{\alpha\alpha}\cos(2k) \right] ~, \\
    J_{2,k} &= 4Jxy\delta a_k a_{k+\pi} \cos(2k) ~, \\
    C_k &= 4 D_1^c a_k^2 \sin(k) ~, \\
    F_k^{\pm\pm} &= 4 a_k a_{k+\pi} \left( \pm \Gamma_1^{ab} \sin(k) \pm D_2^c \sin(2k) \right)
    ~.
\end{align}
\end{subequations}
Upon reproducing the derivation,
we found different signs than in Ref.\ \cite{splinter2016minimal}
in some terms in $\mat{M}^{ab}$ and $\mat{M}^{ba}$.
With the corrected signs the Bogoliubov transformation
yields the pairwise opposite real eigenvalues as required.
We find that the previous calculations underestimate
the energy of the lowest eigenmode.
This has been corrected in our calculations.

We only consider magnetic fields parallel
to one of the three directions $\{a,b,c\}$,
so the submatrices $\mat{H}^{\alpha \beta}$
only mix the triplon operators with flavor
of the two perpendicular directions.
In particular,
if the magnetic field points in $c$ direction,
the matrix is block-diagonal with a 4$\times$4 $c$ block
and a 8$\times$8 $ab$-block.
In this case, the eigenvectors of the $c$ block
are completely independent from $B=|\vc{B}|$.

\subsection{Results}
 
\begin{figure}[t]
 \centering
 \includegraphics[width=1.0\linewidth]{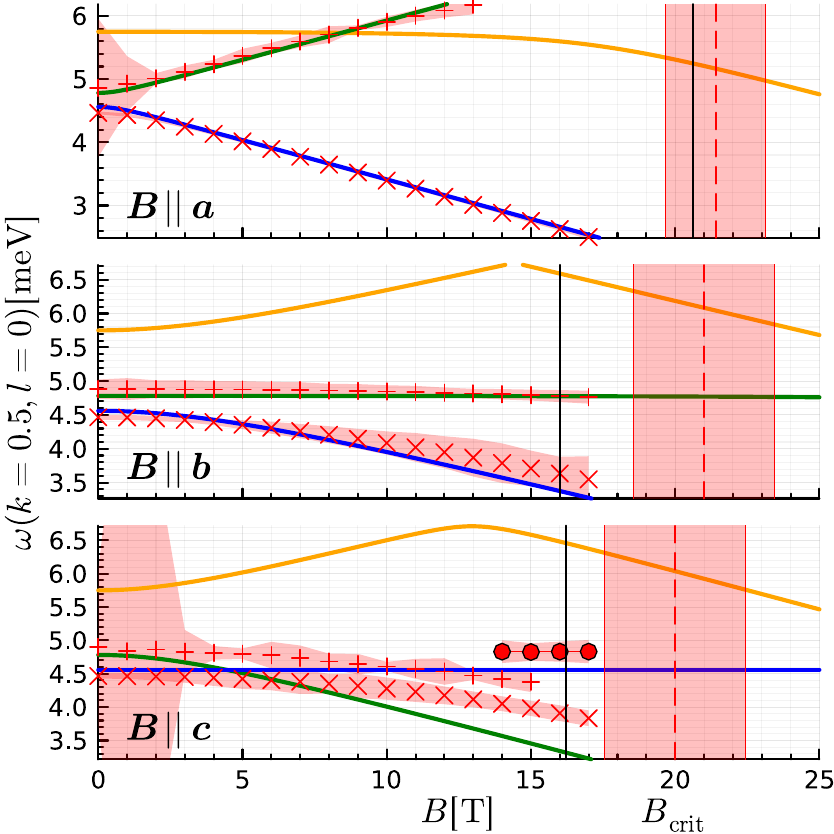}
 \caption{\label{fig:fit}
 Results of minimizing the cost function \eqref{eq:C_full}.
 The three panels  show the fit of the theoretical model energies
 (as blue, green and orange lines)
 to the experimental target data (red crosses with red ribbons indicating the region where the cost function increases by half a unit given the experimental uncertainties and the used weight factors, see also main text)
 for $\vc{B} \parallel \{\vc{a},\vc{b},\vc{c}\}$, respectively.
 The vertical, red line indicates
 the target critical field with corresponding cost regime
 and the vertical, dashed, black line
 the theoretical critical field of the model where the spin gap closes.
}
\end{figure}

The optimization yields
several local minima of the cost function
in the high-dimensional optimization space, for details see App.~\ref{app: Optimization model} where the cost function is defined.
We present the best fits
to the experimental data
in Fig.\ \ref{fig:fit}.
The fit parameters
rounded to the leading digits
are
$(x,y) = (1.2, 0.85)$,
$J = 9.524 \,$meV,
$\lambda = 0.27$,
$\delta = 0.095$,
$\vc{g}=(2.167,1.805,2.0)$,
$D_0^b=0.46$,
$\vc{D}_1=(-0.05, -0.03, -0.02)$
and $(D_2^a, D_2^c)=(-0.06, -0.35)$.
They were obtained in a minimization
with cost parameters
($c_\mathrm{M1,\,M2}$, $c_{\mathrm{M3}, \vc{B} \parallel c}$, $c_\text{gap}$, $c_\text{crit}$) = (1, 1, $10^4$, $5\cdot10^3$)
and $\vc{c}_\text{crit} = (2,1,1)$.

The colors for the modes are chosen
blue, green and orange
for the first, second and third
smallest eigenenergies.
For $\vc{B} \parallel \vc{a}$, we observe that
the second and third lowest mode typically cross.
The algorithm has been implemented
to consider this by reordering
the eigenvalues such that
the modes are kink-free.

The red ribbons visualize the experimental uncertainty,
scaled such that when a mode or data point lies exactly at the edge of the corresponding ribbon,
it adds 0.5 to the total cost.
Therefore, it is a good indicator of how important each data point is for the fitting routine
and of the relative uncertainties of the experimental mode data,
but it should not be mistaken as the absolute experimental uncertainties.
These ribbons also visualizes why the weight factors $c_\text{gap}$ and $c_\text{crit}$ need to be so large:
otherwise, the gap cost and critical cost would be insignificant for the optimization routine, i.e., the corresponding ribbons would be enormously wide.

The fitted modes shows quantitatively very good agreement
for $\vc{B} \parallel \vc{a}$. For $\vc{B} \parallel \vc{b}$ quantitative agreement within the experimental error bars can be achieved. The magnetically hard direction $\vc{B} \parallel \vc{c}$ can capture the correct behavior for small magnetic fields but overestimates the down-bending  of the lowest mode. Interestingly, our theoretical model seems to indicate that the mode $M_3$ (red circles) does not correspond to the third triplon mode, but further studies are called for clarification. 

It is possible to find parameters yielding good fits for $\vc{B} \parallel \vc{c}$ 
by strongly increasing the corresponding weight coefficients in the cost function.
But this comes at the expense of significantly less convincing
agreement of the other data points as well as critical fields.

The found DM interactions are quite large,
in particular $D_0^b=0.46$.
It must be noted that since triplon-triplon interactions
are not taken into account in the employed bilinear model,
the fit parameters are effective DM 
interactions that are known to be larger
than the ones when taking triplon interactions into account \cite{mueller2023interacting, plumb2016quasiparticle}.
The same applies to the \textit{g} factors, which are in the range $1.8-2.2$. They have to be interpreted as effective values within the bilinear model and may change due to renormalization effects if the triplon-triplon interaction is taken into account.

Comparing the exchange and DM interactions to the single particle bond-operator theory from Ref.~\cite{plumb2016quasiparticle}, we observe that our CUT and systematic fitting procedure yields a different set of DM couplings, which are slightly smaller, as well as a  modified set of isotropic exchange parameters. 
Using the notation from Ref.~\cite{plumb2016quasiparticle}, they employed $J_{1}=J_{2}=J_{4}=8$~meV and found a largest DM interaction of $D_{1}^{a}=0.6J_{4}$, whereas our fit favors $J_{4}=9.524$~meV, $J_{1}=1.2J_{4}, J_{2}=0.85J_{4}$, and the largest DM interaction at $D_{4}^{b}=0.46J_{4}$. We attribute these discrepancies to the different theoretical approaches as well as the availability of multiple parameter sets capable of describing the low-energy features.
Notably, a small discrepancy of about $1.5$~meV between the THz data at the $\Gamma$ point and the INS data at $k=0.5$ falls within the experimental uncertainties of the INS measurements. Our model accurately reproduces the INS gap at $k=0.575$ at zero field and captures both the zero- and finite-field THz modes, indicating that, within these uncertainties, the THz and INS experiments are consistent with each other.

\section{Conclusion}

In summary, we document a comprehensive terahertz spectroscopy study of low-energy spin dynamics in a low-dimensional quantum magnet BiCu$_2$PO$_6$ at temperatures down to 1.4~K and in magnetic fields up to 60~T.
For the magnetic fields applied in different crystallographic orientations, we observed evidently different field dependencies of the spin excitations.
In particular, for field along the $a$ axis, a field-induced phase transition at $B_{c1}=21.4$~T is evidenced by the observation of a low-energy excitation above $B_{c1}$.
The clear difference in the field-dependent evolution of the spin excitations indicates a strong magnetic anisotropy of the spin interactions.

To take into account the magnetic anisotropies theoretically, we introduce bond-dependent Dzyaloshinskii-Moriya interactions as well as the concomitant symmetric anisotroic exchanges and anisotropic \textit{g} tensors in our model Hamiltonian, in addition to the Heisenberg exchange interactions between nearest- and next-nearest-neighbor spins.
We performed a largely unbiased and automatic analysis of the magnetic dispersions in BiCu$_2$PO$_6$ on the level of a bilinear model with focus on the low-lying modes, by defining suitable cost functions that include the expected dispersion minimum and the critical magnetic field in a computationally efficient manner.
With the determined fits, we observe very good agreement with the spectroscopic results in both the $a$ and $b$ field directions. While the low-field results in the $c$ direction also align well, our analysis indicates that the theoretical model overestimates the downward bending of the lowest mode. Additionally, our model rules out the possibility of the high-field $M_3$ mode being the third triplon, raising questions about the physical origin of this mode. 

From the theory side, it is desirable to investigate the spin dynamics by an explicit inclusion of triplon-triplon interactions in the presence of magnetic fields. In zero field, calculations of this kind can be found in Refs.~\cite{plumb2016quasiparticle,mueller2023interacting}. 
From the experimental side, a comprehensive measurement of the spin excitations and possible excitation continua in magnetic fields by inelastic neutron scattering with energy and momentum resolution is highly interesting.
For field applied along other directions than the \textit{b} axis, a nuclear magnetic resonance measurement may reveal other features of field-induced phases than soliton lattice \cite{casola2013DirectObservation}.

\section*{acknowledgments}
We thank Y. Kohama for helpful discussions, and acknowledge support by the European Research Council (ERC) under the Horizon 2020 research and innovation programme, Grant Agreement No. 950560 (DynaQuanta), by the Estonian Ministry of Education, personal research funding PRG736, and by the European Regional Development Fund project TK134 as well as by the Deutsche Forschungsgemeinschaft (German Science Foundation) in project UH90-14/1. The authors wish to thank the National Science and Technology Council for their support of this project under contract NSC111-2112-M-027-004-MY3.
This work was supported by the Deutsche Forschungsgemeinschaft through the W\"{u}rzburg-Dresden Cluster of Excellence on Complexity and Topology in Quantum Matter - $ct.qmat$ (EXC 2147, project No. 390858490) and the SFB 1143, as well as by the HFML-RU/FOM and the HLD at HZDR, members of the European Magnetic Field Laboratory (EMFL).

\bibliography{BCPO_bib}

\appendix

\section{Optimization of Model Parameters}
\label{app: Optimization model}
We perform a numerical optimization to
fit the model to the experimental data
with as little bias as possible.
First, we perform the CUT for a fixed pair of relative ladder couplings $(x,y)$.
Ref.\ \cite{splinter2016minimal} analyzed
which pairs $(x,y)$ yield a reasonable dispersion and found 
a good approximation for $(x_0,y_0)=(1.2, 0.85)$.
For our analysis, we performed computations with other pairs in the vicinity of $(x_0,y_0)$ as well, but the pair $(1.2, 0.85)$ yields the best results.
For any given pair $(x, y)$,
the other parameters $\vc{\theta} 
= (J, \lambda, D_0^b, \vc{D}_1, D_2^a, D_2^c, \vc{g}, \delta)$ are numerically optimized,
using the diagonalization of the matrix $\mat{M}(\vc{k},\vc{B})$
in order to compute the magnetic dispersions $\omega_{m_n}^\alpha(\vc{k},B^\alpha_n)$,
where $\alpha\in\{a,b,c\}$ denotes the direction of the magnetic field $\vc{B}_n \parallel \vc{e}_\alpha$.
The index $n$ stands for the concrete experimental data point for mode $m_n$ and magnetic field $B_n^\alpha$.

We optimize the parameters in such a way that they minimize a suitably chosen 
cost function in Eq.~(\ref{eq: cost function}).
This cost function consists of several contributions. The term
$C_{M1,M2}$ is the least-square sum of the deviations of the modes $M_1$ and $M_2$; similarly
$C_{M3, \vc{B}\parallel \vc{c}}$ measures the deviations of the third mode if the magnetic field is applied along  the $c$ direction.
The partial cost $C_\text{gap}$ captures the deviation in the spin gap while $C_\text{crit}$
assesses the deviation in the critical magnetic fields at which the valence bond phase breaks down.
In total, we consider
\begin{widetext}
\begin{subequations}
\begin{align}
    C(\vc{\theta}) &= \frac{C_\mathrm{M1,\,M2}(\vc{\theta}) + C_{\mathrm{M3}, H\parallel z}(\vc{\theta}) + C_\text{gap}(\vc{\theta}) + C_\text{crit}(\vc{\theta})}{c_\text{all}} ~, \label{eq:C_full} \\
    C_{M1,M2}(\vc{\theta}) &= \frac{c_{M1,M2}}{N_{\mathrm{M1,\,M2}}} \sum_{n} \left| \frac{\omega_{m_n}^{\alpha}(\vc{\theta}, \vc{k}=(\frac{\pi}{2}, 0), B_n^\alpha)-\omega_{m_n}^{\alpha\text{, target}}}{ \Delta_n^\alpha } \right|^2 ~, \\
    C_{M3, H\parallel z}(\vc{\theta}) &= c_{\mathrm{M3}, \vc{B}\parallel \vc{c}} \left| \frac{ \omega_{\mathrm{M3}, \vc{B}\parallel \vc{c}}(\vc{\theta},\vc{k}=(\frac{\pi}{2}, 0), B_n^\alpha) - \omega_{\mathrm{M3}, \vc{B}\parallel \vc{c}}^\text{target} }{ \Delta_{\mathrm{M3}} } \right|^2 ~, \\
    C_\text{gap}(\vc{\theta}) &= c_\text{gap} \left( \left| \frac{|k_\text{gap}(\vc{\theta}) - k_\text{gap}^\text{target}}{k_\text{gap}^\text{target}}\right|^2 + \left| \frac{\omega_\text{gap}(\vc{\theta}) - \omega_\text{gap}^\text{target}}{\omega_\text{gap}^\text{target}} \right|^2 \right) ~, \\
    C_\text{crit}(\vc{\theta}) &= \frac{c_\text{crit}}{3} \sum_\alpha c_\text{crit}^\alpha \bigg| \omega_\text{gap}^\alpha(\vc{\theta}, k^\alpha_\text{crit}, B_\text{crit}^{\alpha,\text{target}}) \bigg|^2 \bigg/ \sum_\alpha c_{\mathrm{crit}}^\alpha ~, \label{eq:C_crit} \\
    k^\alpha_\text{crit} &= \underset{k}{\text{minarg}} \bigg(~ \bigg| \real\, \omega_\text{gap}^\alpha(\vc{\theta},k, B_\text{crit}^{\alpha,\text{target}}) \bigg| - \bigg| \imag\, \omega_\text{gap}^\alpha(\vc{\theta},k, B_\text{crit}^{\alpha,\text{target}}) \bigg| ~\bigg) ~,\\
    c_\text{all} &=  c_\mathrm{M1,\,M2} + c_{\mathrm{M3}, \vc{B}\parallel \vc{c}} + c_\text{gap} + c_\text{crit} ~.
\end{align}
\label{eq: cost function}
\end{subequations} 
\end{widetext}
The coefficients ($c_\mathrm{M1,\,M2}$, $c_{\mathrm{M3}, \vc{B} \parallel \vc{c}}$, $c_\text{gap}$, $c_\text{crit}$) are weight factors,
which are tuned by hand in order to find good overall fits,
and $N_{\mathrm{M1},\,\mathrm{M2}}$ is the number of sampling points of the dispersions used in the fit.
The sampling points we use here are the terahertz spectroscopy results reported in this work.
For the presented data, the weights were chosen to be (1, 1, $10^4$, $5\cdot10^3$) and $\vc{c}_\text{crit} = (2,1,1)$.
Note that due to the differences in how the cost terms are computed,
setting $c_\mathrm{gap}= 10 c_\mathrm{M1,\,M2}$ does not mean that
the error in the gap is 10 times more relevant for the optimization
than other errors.

The $M_3$ mode for $\vc{B} \parallel \vc{c}$ is treated separately in $C_{\mathrm{M3}, \vc{B} \parallel \vc{c}}(\vc{\theta})$
instead of incorporating it in $C_{\mathrm{M1,\,M2}}(\vc{\theta})$
because the experimental data suggests that it is a constant, flat mode.
Therefore, we fit it only to the constant modes obtained from the 4$\times$4 $c$ block.

The experimental uncertainties $\Delta_n^\alpha$
for the experimental data of the modes $M_1$ and $M_2$
are used to weight the fits according to the accuracy of the data.
For the $M_3$ mode, the geometrically averaged  $\Delta$ of all data points is used
because the fitted mode is constant anyway.

The term $C_\text{crit}(\vc{\theta})$
quantifies the deviation in the
critical magnetic field.

\end{document}